\documentclass[sigconf]{acmart}
\AtBeginDocument{%
  \providecommand\BibTeX{{%
    \normalfont B\kern-0.5em{\scshape i\kern-0.25em b}\kern-0.8em\TeX}}}
    
\usepackage{tabularx}
\usepackage{hyperref}

\copyrightyear{2022}
\acmYear{2022}
\setcopyright{rightsretained} 
\acmConference[CHI '22 Extended Abstracts]{CHI Conference on Human Factors in Computing Systems Extended Abstracts}{April 29-May 5, 2022}{New Orleans, LA, USA}
\acmBooktitle{CHI Conference on Human Factors in Computing Systems Extended Abstracts (CHI '22 Extended Abstracts), April 29-May 5, 2022, New Orleans, LA, USA}
\acmDOI{10.1145/3491101.3516800}
\acmISBN{978-1-4503-9156-6/22/04}

\begin{document}

\title{The Effect of Multiple Replies for Natural Language Generation Chatbots}

\author{Eason Chen}
\orcid{0000-0003-1486-8559}
\affiliation{%
  \institution{National Taiwan Normal University}
  \city{Taipei}
  \country{Taiwan}
}
\email{eason.tw.chen@gmail.com}


\begin{abstract}
In this research, by responding to users’ utterances with multiple replies to create a group chat atmosphere, we alleviate the problem that Natural Language Generation chatbots might reply with inappropriate content, thus causing a bad user experience. Because according to our findings, users tend to pay attention to appropriate replies and ignore inappropriate replies. We conducted a 2 (single reply vs. five replies) $\times$ 2 (anonymous avatar vs. anime avatar) repeated measures experiment to compare the chatting experience in different conditions. The result shows that users will have a better chatting experience when receiving multiple replies at once from the NLG model compared to the single reply. Furthermore, according to the effect size of our result, to improve the chatting experience for NLG chatbots which is single reply and anonymous avatar, providing five replies will have more benefits than setting an anime avatar.
\end{abstract}



\begin{teaserfigure}
    \centering
    \includegraphics[width=\linewidth]{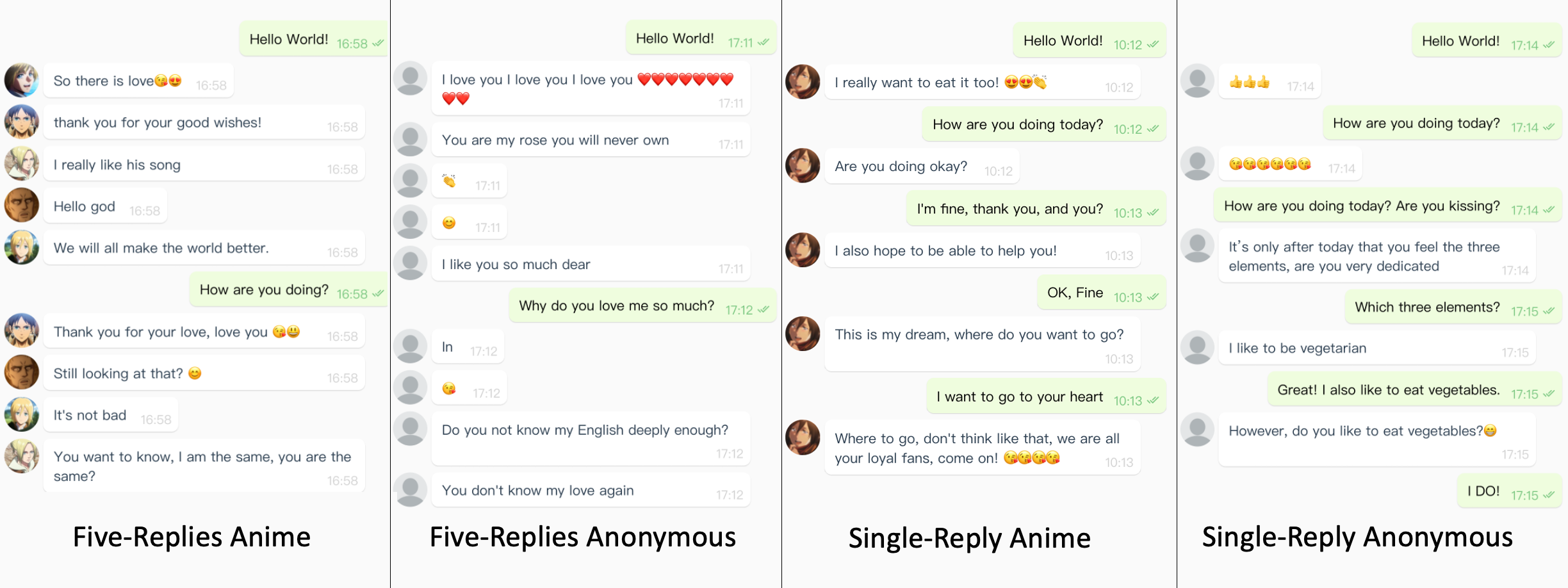}
    \caption{The Screenshot of 2 (single reply vs. five replies) $\times$ 2 (anonymous avatar vs. anime avatar) experimental conditions. From left to right is Five-Replies Anime, Five-Replies Anonymous, Single-Reply Anime, Single-Reply Anonymous, respectively.}
    \label{fig:fig1}
\end{teaserfigure}

\maketitle

\section{Introduction}
``Chatbot can be a software which interacts with people using natural language by building a conversation'' \cite{gunasekara2019uniontbot}. Chat-oriented chatbots are becoming popular in providing chatting services \cite{gunasekara2019uniontbot} thanks to the recently developed neural language modeling \cite{bengio2000neural}, especially Natural Language Generation model, like GPT-2 \cite{Radford2019LanguageMA}.

To generate text, Natural Language Generation (NLG) model adopts a decoding strategy, such as the top-k or Nucleus method \cite{holtzman2019curious}, to generate one token at a time based on the input text and currently generated tokens until a certain stop criterion is met. Due to the randomness nature of the decoding strategy \cite{holtzman2019curious}, the generated token sequence will be different even when the same input sequence is given. Therefore, it is hard to predict the quality of a single output based on the input. Consequently, during the conversation between the user and the chatbot, especially on open-domain topics, the current NLG chatbots inevitably make errors \cite{higashinaka2015fatal} and affect the chatting experience.

We notice that most NLG chatbot researches focus on providing a single reply which is the best output by the best model \cite{gunasekara2019uniontbot, yang2020development, rahman2017programming}. However, since the output from the NLG model can't be perfect, we consider another approach to let users maintain a satisfactory chatting experience:  providing more replies. In this research, we want to know how multiple replies influence the chatting experience for NLG Chatbot. Therefore, we recruited 102 participants to join a 2 (single reply vs. five replies) $\times$ 2 (anonymous avatar vs. anime avatar) repeated-measures experiment and analyzed with two-way ANOVA to compare the chatting experience in different conditions (see \autoref{fig:fig1}).

In sum, the main contributions for this paper are:
\begin{enumerate}
\item 	According to our survey, we find users tend to pay attention to appropriate replies and ignore inappropriate replies when chatting with the chatbot.
\item 	Compared to a chatbot with single reply, the result from our experiment shows that users have a better chatting experience when the chatbot responds to users’ one utterance with multiple replies.
\item 	According to the effect size of our result, to improve the chatting experience for an NLG chatbot which is single reply and anonymous avatar, providing multiple replies will have more benefits than setting an anime avatar.
\end{enumerate}

\section{Theoretical Framework and Hypotheses}
When  using a chat-oriented chatbot, the key to having a good chatting experience is to avoid dialog breakdown. Dialog breakdown means users are feeling difficult to continue their conversation \cite{higashinaka2015fatal}. The main reason causing dialog breakdown is inappropriate messages, which includes ignoring the user’s input, providing a not understandable reply, repetition, or contradictory content \cite{higashinaka2015fatal}. Therefore, we suppose the frequency of suitable and unsuitable messages can predict the chatting experience.
\textit{H1: The frequency of appropriate replies can significantly predict the chatting experience score.}

Because the NLG model’s output is highly randomized \cite{holtzman2019curious}, it is hard to maintain the user’s chatting experience. Once users receive an unacceptable answer, they might stop chatting. However, we can view it differently: users will continue chatting if they receive an acceptable reply. So, how to increase the likelihood that users receive a proper response? We can increase the number of replies. In purely rational analysis, this is a question of probability: assuming that given the user’s input x, the probability that the model generates a suitable response will be p(x) (ex: 0.4). Then the probability that the model can’t generate a suitable response will be 1 – p(x) (ex: 0.6). However, if the model generates five parallel responses at once, the probability that the model can’t generate any fitting response will be (1 – p(x))5 (ex: 0.65 = 0.0776). Compared with the single reply, the chance that the user will not receive any proper reply in the multi reply condition will be greatly reduced (ex: from 0.6 to 0.0776). In other words, with more replies, the probability that the model will generate at least one suitable answer will be greatly increased (ex: from 0.4 to 0.9224). Thus, users might have a better chatting experience in the multi replies condition than the single reply condition. This leads to the following hypothesis:
\textit{H2: Multiple replies condition has a better user experience than the single reply condition.}

The hypothesis  mentioned above be valid or not highly dependent on users’ mental processes. According to the Sensemaking theory \cite{weick2005organizing}, people will interpret the information they receive in a way that they can understand. ``Explicit efforts at sensemaking tend to occur when the current state of the world is perceived to be different from the expected state of the world'' \cite{weick2005organizing}. Hence, users will be trying to make sense of what the chatbot wants to say when interacting. Accordingly, suppose we present more replies to create a group chat atmosphere. Then, it might be easier for users to comprehend what the chatbot is saying and increase tolerance for unfitting responses. For example, users might not take those unfitting messages seriously since they assume the chatbot may not be talking to them but to other chatbots in the same chatroom. Therefore, we propose the following hypothesis:
\textit{H3: Users tend to pay attention to appropriate replies and ignore inappropriate replies when interacting with chatbots.}

\begin{figure}[t]
    \centering
    \includegraphics[width=\linewidth]{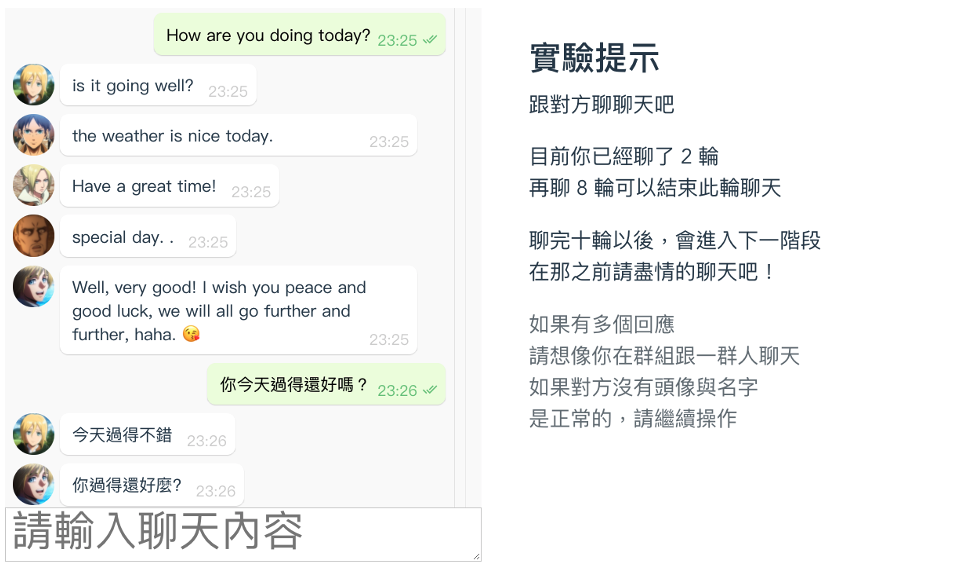}
    \caption{The Screenshot of the experimental platform at five-replies anime condition.  The left is the chatroom, and the right is the instruction of the test.}
    \label{fig:fig2}
\end{figure}

Nonetheless, when providing multiple replies, we received feedback from pre-study participants who said that it is hard to distinguish the replies as ‘many from one person’ or ‘many from many people’ without avatar. Therefore, we provide avatars for each reply as a control condition to help participants differentiate the multiple replies. We suppose that at the chatting experience score, there is an interaction effect between the avatar type and the number of replies.
\textit{H4: There is an interaction effect between avatar type and the number of replies at chatting experience score.}

\section{Research Method}
The experiment used a 2 (single reply vs. five replies) $\times$ 2 (anonymous avatar vs. anime avatar) repeated measures design. Each participant was required to undergo four randomly ordered experimental conditions. Apart from that, we choose five replies here because it is best to fit the chatroom window without overflow while maximizing the replies. And the anime avatars for chatbots are selected from the popular anime “Attack on Titan” (see in \autoref{fig:fig1} and \ref{fig:fig2}).

\subsection{Participants and Procedure}
We recruited 102 participants in Taiwan from the internet through convenience sampling with a mean age of 22.3 years (range 19 – 50, SD = 3.89). At the experiment, participants will first login to the platform with their own computer and read the instruction of the experiment. Following the instruction, participants will have ten times of chat warmup to get familiar with the platform. After that, participants will go through one of the four conditions. Each condition participant will first chat ten rounds (\autoref{fig:fig2}) then fill the form for the chat experience of the round. After all four rounds are finished, the participant will fill the final form for suitable response frequency and their attention tendency. 

\subsection{Measures}
All questionnaires are measured by Google Form.

\subsubsection{Chatting  Experience}
We translated The Chatbot Usability Questionnaire (CUQ) \cite{holmes2019usability} into Chinese to measure the chatting experience in each round. The scale is designed to measure the user experience of using chatbots. The participant must select 1 (strongly disagree) to 5 (strongly agree) for sixteen statements (such as The chatbot understood me well, or Chatbot responses were irrelevant). According to the author \cite{holmes2019usability}, the score of CUQ has a significant positive correlation with the score from System Usability Scale (SUS) \cite{brooke1996sus} and the User Experience Questionnaire (UEQ) \cite{laugwitz2008construction}. Furthermore, based on the data we collected, the value for Cronbach’s Alpha for the CUQ was $\alpha = .94$.

\subsubsection{The frequency of appropriate replies}
To measure the frequency of appropriate replies, we use a custom five-point scale (\autoref{tab:tab1}). Participants must select 1 (strongly disagree) to 5 (strongly agree) for the following two statements. The frequency of suitable replies will be obtained by the score of the first question, plus six minus the score of the second question. These statements were verified with three experts from relative fields, and based on the data we collected, the value for Cronbach’s Alpha for this survey was $\alpha = .83$. Participants will review the definition of appropriate and inappropriate replies before measurement begins.

\begin{table}[t]
    \caption{Questions to measure the frequency of suitable replies.}
    \label{tab:tab1}
    \centering
    \begin{tabularx}{\linewidth}{lX}
    \toprule
    \textbf{Nr.} & \textbf{Statement}\\
    \midrule
1	& In the previous chat, there are many ‘appropriate’ replies\\
2	& In the previous chat, there are many ‘inappropriate’ replies\\
    \bottomrule
    \end{tabularx}
\end{table}

\subsubsection{Attention/Ignore tendency for appropriate replies and inappropriate replies}
To measure participants’ attention tendency, we use a custom five-point scale (\autoref{tab:tab2}). The participants must select 1 (strongly disagree) to 5 (strongly agree) for the following four statements. These statements were verified with three experts from relative fields, and based on the data we collected, the value for Cronbach’s Alpha for this survey was 
$\alpha = .81$. To compare users’ attention tendency, we will use the paired samples t-test for questions 1, 2, and 3, 4, respectively. 

\begin{table}[t]
    \caption{Questions to measure attention/ignore tendency for appropriate replies and inappropriate replies.}
    \label{tab:tab2}
    \centering
    \begin{tabularx}{\linewidth}{lX}
    \toprule
    \textbf{Nr.} & \textbf{Statement}\\
    \midrule
1 &	In the previous chat, I often pay attention to the appropriate replies \\
2 &	In the previous chat, I often pay attention to inappropriate replies \\
3 &	In the previous chat, I often ignore appropriate replies \\
4 &	In the previous chat, I often ignore inappropriate replies \\

    \bottomrule
    \end{tabularx}
\end{table}

\subsection{Experimental Platform}
The experimental platform is a WEB service with front-end and back-end. The front-end build using Vue.js and Bootstrap. The back-end uses Python FastAPI with MongoDB to process the data. We connect Google Sheet API with the platform so participants can fill the questionnaire at Google Form, and the platform will check their progress via Sheet API to make sure they finish and can go to the next round. The platform's source code is available at GitHub\footnote{EasonC13, AI\_Chatbot\_experiment\_backend: \url{https://github.com/EasonC13/AI_Chatbot_experiment_backend}}.

\begin{table*}[t]
    \caption{The regression result for the frequency of suitable replies predicts CUQ score of the four groups. (Note: *** p < .001).}
    \label{tab:tab3}
    \centering
    \begin{tabularx}{\linewidth}{X*{8}{c}}
    \toprule
    & \multicolumn{2}{l}{\textbf{Five-Replies-Anime}} & \multicolumn{2}{l}{\textbf{Five-Replies Anonymous}} & \multicolumn{2}{l}{\textbf{Single-Reply Anime}} & \multicolumn{2}{l}{\textbf{Single-Reply Anonymous}}\\
    			
    \midrule
The frequency of suitable replies & .576 & 7.046*** & .514 & 5.997*** & .339 & 3.607*** & .360 & 3.860*** \\
F & 49.646*** & 35.969*** &  & 13.014*** &  & 14.897***  \\
Adjusted R2 & .325 & .257 &  & .106 &  & .121 \\
    \bottomrule
    \end{tabularx}
\end{table*}

\subsection{NLG Model}
About the NLG model, we use the design from Yang and Tseng \cite{yang2020development}, which is a GPT-2 and BERT dual-model reply generation system. To be specific, after receiving the input message, GPT-2 will first generate many replies and use BERT to predict replies’ coherence score. Then, the system will sort the coherence score and use top n replies as the output. The system can successfully input various dialogues and adjust emotions to output appropriate responses after learning about the 1.7 million corpus provided by the Japanese NTCIR Chinese Emotional Conversation Generation (CECG) evaluation task in 2019 \cite{zhang2019overview}. To control the variance in the experiment, we only use ‘like’ as the emotion type and generate one response per request. Moreover, even though the model is Chinese-based, we connect the Google Translate API so that the chat system can reply according to the detected language. That is why some texts in screenshots are English.

\section{Result}
The data analysis was performed using SPSS 23.0.

\subsection{Attention/Ignore tendency for appropriate replies and inappropriate replies}
\begin{figure}[t]
    \centering
    \includegraphics[width=\linewidth]{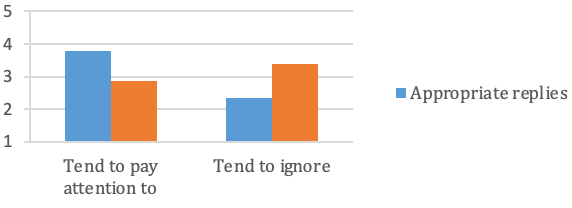}
    \caption{Attention/Ignore tendency survey result for appropriate replies and inappropriate replies}
    \label{fig:fig3}
\end{figure}

The result from the attention/ignore tendency (\autoref{fig:fig3}) showed that the participants tend to pay attention to the appropriate replies ($M = 3.78$, $\textit{SD} = 1.07$) more than inappropriate replies ($M = 2.88$, $\textit{SD} = 1.14$) ($t(97) = 4.521$, $p < .001$, $d = 0.457$) and ignore inappropriate replies ($M = 2.33$, $\textit{SD} = 0.95$) more than appropriate replies ($M = 3.40$, $\textit{SD} = 1.182$) ($t(97) = -5.890$, $p < .001$, $d = 0.595$). \textbf{H3 Accepted}.

\subsection{The relation between the frequency of suitable replies and the CUQ Score}
Four simple linear regressions were carried out to test if the frequency of suitable replies ($M = 3.88$, $\textit{SD} = 1.58$) predicted the CUQ score of the four groups. All four results of the regressions are significant (\autoref{tab:tab3}). Indicated that the frequency of suitable replies significantly predicted chatting experience. Moreover, the result shows that the frequency of suitable replies can predict the CUQ score more on five-replies conditions. (H1 Accepted)

\subsection{Comparison of the CUQ scores in the four groups }
\begin{figure}[t]
    \centering
    \includegraphics[width=\linewidth]{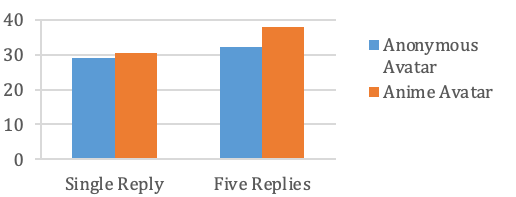}
    \caption{The difference between the four groups.}
    \label{fig:fig4}
\end{figure}

CUQ scores are subjected to a two-way analysis of variance for the following four conditions: 2 (single reply vs. five replies) $\times$ 2 (anonymous avatar vs. anime avatar). The interaction effect for CUQ score between reply count and avatar type is non-significant ($F(1, 97) = 2.13$, $p = .148$). Therefore, H4 can be rejected. Regards to the CUQ scores for the four groups (\autoref{fig:fig4}), from high to low are Five-Replies-Anime ($M = 37.5$, $\textit{SD} = 20$), Five-Replies Anonymous ($M = 30.9$, $\textit{SD} = 18.4$), Single-Reply Anime ($M = 30.5$, $\textit{SD} = 18.8$), Single-Reply Anonymous ($M = 29.0$, $\textit{SD} = 20.1$).
Furthermore, the main effect of the number of replies indicates that for CUQ score, the five replies conditions are significantly higher than the single reply conditions ($F(1, 97) = 10.09$, $p = .002$, $\eta^2 = .094$). (H2 Accepted)
Apart from that, the main effect of avatar type indicates that for CUQ score, the anime avatar conditions are significantly higher than the anonymous avatar conditions ($F(1, 97) = 4.57$, $p = .035$, $\eta^2 = .045$).

\section{General Discussion}
The result indicates the following:
1. When users chat with an NLG Chatbot, they tend to pay attention to appropriate replies and ignore inappropriate replies.
2. When designing an NLG chatbot, if responding to a user’s utterance with multiple replies, simulating a multi-person group chat will bring a better chatting experience. Therefore, it is easier to satisfy users with multiple responses, which can reduce the influence of occasional poor answers of the system.
3. There is no interaction effect between the avatar and the number of replies on the chatting experience. Thus, increasing the number of answers while providing multiple virtual avatars can bring a higher chatting experience score. Also, according to the effect size ($\eta^2$), to improve the chatting experience for an NLG chatbot which is single reply and anonymous avatar, providing multiple replies ($\eta^2 = .094$) might benefit more than setting an anime avatar ($\eta^2 = .045$).
Our finding demonstrates the potential of letting chatbot to response to user’s utterance with multiple replies. At present, most NLG chat-oriented chatbots only provide one reply. Therefore, we hope that developers can refer to this research and put multiple replies into design considerations when developing a chat-oriented chatbot in the future.

\begin{acks}
This work was supported by the Ministry of Science and Technology of Taiwan (R.O.C.) under Grants 110-2813-C-003-033-E. We thank Yuen-Hsien Tseng, Tzung-Jin Lin, and Ching-Lin Wu for commenting on our manuscript and suggestion from Guo-Li Chiou, Chien Wen Tina Yuan, Tsung-Ren Huang and Liang-Yi Li.
\end{acks}

\bibliographystyle{ACM-Reference-Format}
\bibliography{bibliography}

\end{document}